\documentstyle[prl,aps]{revtex}
\twocolumn
\begin{document}
\input{epsf}
\draft
\twocolumn[\hsize\textwidth\columnwidth\hsize\csname@twocolumnfalse\endcsname
\title {Weak selection and stability of 
localized distributions in Ostwald ripening}
\author{Boaz Giron, Baruch Meerson and Pavel V. Sasorov \cite{adr}}
\address{The Racah Institute  of  Physics, Hebrew
University   of  Jerusalem,
Jerusalem 91904, Israel}
\maketitle
\begin{abstract}
We support and generalize a weak selection rule
predicted recently for the self-similar asymptotics 
of the
distribution function (DF) 
in the zero-volume-fraction limit of Ostwald ripening 
(OR). An asymptotic perturbation theory is developed that, when 
combined with an
exact invariance property of the system, yields the selection rule, 
predicts
a power-law convergence towards
the selected self-similar DF
and agrees well with our numerical  simulations for the interface- 
and diffusion-controlled OR.
\end{abstract}
\vskip1pc]
\narrowtext
  
In a late stage of a first-order phase transition
a two-phase mixture undergoes coarsening, or Ostwald ripening (OR), when 
the minority phase tends to minimize its interfacial energy under condition 
of
a constant volume \cite{LS1,LS2,W}. Despite
numerous works, OR continues to attract attention
both in experiment \cite{experiment} and in 
theory \cite{theory,MS96}. Our main motivation in studying this problem
has been an attempt to resolve an old
selection problem (described below) that created much controversy. 

The 
``classical" formulation of the problem of OR, valid in the limit of a 
negligibly small volume fraction of the 
minority domains,
is due to
Lifshitz and Slyozov (LS) \cite{LS1,LS2}  and 
Wagner \cite{W}. 
In this formulation, the dynamics  
of the distribution function (DF) 
$F(R,t)$  of the domain sizes 
are governed (in scaled variables) by the continuity
equation
\begin{equation} 
\frac{\partial F}{\partial t}+\frac{\partial}{\partial R}
(V F)=0\,,\quad 
V (R,t)= \frac{1}{R^n} \left(\frac{1}{R_c}-\frac{1}{R}\right)\,,
\label{1}
\end{equation}
where $R_c (t)$ is the critical radius for expansion/shrinkage of an individual
drop, while $n$ 
is determined by the mass transfer mechanism. 
The dynamics
are constrained by conservation of the total volume of the minority domains:
\begin{equation} 
\int_{0}^{\infty} R^3\,F(R,t)\,dR \,= Q = const \,.
\label{2}
\end{equation}

Of great interest are possible self-similar intermediate asymptotics of
this problem and the rule that selects the relevant asymptotics 
out of many possibilities. Scaling analysis of 
Eqs. (\ref{1}) and (\ref{2}) yields a similarity 
ansatz $F(R,t)=t^{-\mu}\, \Phi \,(R\,t^{-\nu})$ and $R_c= (t/\sigma)^{\nu}$,
where $\mu=4/(n+2)$, 
$\nu = 1/(n+2)$ and $\sigma=const$. Upon substitution, one finds
a {\it family} of self-similar DFs
for every $n \ge -1$, where each of
the DFs is localized on a finite interval 
$[0, u_m]$
of the similarity variable 
$u=R\,t^{-\nu}$. The DFs can be parameterized by 
$\sigma$, and the interval of possible values of $\sigma$ is determined 
by the
requirements of the continuity of $F(R,t)$ on the whole interval
$[0, u_m]$ and normalizability
with respect to Eq. (\ref{2}). For each of the solutions, the
average domain radius grows in time like $t^{1/(n+2)}$, and the number 
of domains decreases like $t^{-3/(n+2)}$, but the coefficients in 
these scaling laws are $\sigma$-dependent.

The self-similarity and related scalings were 
discovered by LS \cite{LS1,LS2} in the case of $n=1$ 
(diffusion-controlled OR).
LS arrived at a {\it unique} self-similar DF  (we 
will call it the limiting solution), and ruled out 
other possible solutions. In the first paper \cite{LS1},  
the other solutions were rejected 
as non-normalizable with respect to Eq. (\ref{2}). 
In the case of $n=0$ (interface-controlled
OR) this 
argument was repeated by Wagner \cite{W}. 
However, already in their second 
paper \cite{LS2} on the same subject 
LS realized that no 
problem with normalization
arises for initially localized DFs (that is, for those
with a compact support 
at $t=0$). 
This correction was apparently 
overlooked in the literature ({\it e.g.}, Ref. \cite{Pitaevsky}), until
Brown \cite{Brown} 
addressed the other solutions 
and found
them numerically for $n=1$. This created a long-standing 
controversy (see, {\it e.g.} 
\cite{Chen}), and the first step towards resolving it was
made in the case of $n=0$ \cite{MS96}.
It was noted that a
DF, initially localized on an interval $[0,\,R_m(t=0)]$,
always remains localized  
on a (time-dependent) interval
$[0,\,R_m(t)]$.
Furthermore, if 
$F(R,\,t=0)$ is describable 
by a power law
$A_0 \,[R_m(t=0) - R]^{\lambda}$ in the close vicinity of $R=R_m(t=0)$, then 
for any $t>0$ the leading term in the expansion of $F(R,\,t)$ in the vicinity
of $R=R_m(t)$ has the form $A (t) \, [R_m(t) - R]^{\lambda}$. 
Invariance of the exponent $\lambda$ under 
the dynamics (\ref{1}) and (\ref{2}) 
implies a (weak) selection rule for the 
``correct" self-similar
DF, as there is a one-to-one correspondence between
$0<\lambda<+\infty$ and the parameter $\sigma$ \cite{MS96}. 
[The limiting solution 
corresponds to an
extended (non-compact) initial condition or, formally, to
$\lambda \rightarrow +\infty$.] More precisely, if a self-similar
asymptotics is ever reached, it must be the one selected by $\lambda$.
However, no
attempts have been made to solve (even numerically) 
the full time-dependent problem with 
a localized initial DF. Furthermore, no
stability/convergence analysis for the
localized DFs has been performed, 
so the selection rule proposed in \cite{MS96} has remained unconfirmed.

This Letter supports the selection rule along three directions. 
The first one is to generalize
the selection rule for any $n \ge -1$.  
The second is to
prove the
stability of and analyze the convergence towards the selected 
self-similar DF. The third is to verify our theoretical predictions 
numerically.

A meaningful formulation of 
the stability problem requires some care. Indeed,
each member of the family of self-similar solutions for the DF, except 
the limiting solution, is formally
unstable with respect to addition of an (infinite) tail. In this case
it is the limiting solution that will finally 
develop \cite{LS1,LS2,Chen}. However, such a perturbation is not
always possible.
In addition, the results of \cite{MS96} imply that each
member of the family, except the limiting solution, 
is formally unstable with respect to a
{\it localized} 
perturbation that either has a larger $R_m$ than the ``unperturbed" DF, 
or the same $R_m$  and a 
smaller exponent
$\lambda$.  In each of these cases another 
self-similar solution
from the same family finally develops (as we see in our 
numerical simulations), and 
this situation can hardly be
regarded as instability. A meaningful formulation of the stability problem
should therefore deal with initial
perturbations
localized on the same interval of $R$  as
the ``unperturbed" DF, and characterized by the same exponent
in the close vicinity of $R_{m}(t=0)$. 

We will develop an asymptotic 
linear theory that, combined
with an (exact) invariance property of the model, will enable us to
prove, analytically, the stability of each of the self-similar DFs. 
This result and our numerical simulations will confirm 
the weak selection rule 
\cite{MS96}. We will analyze 
the late-time convergence of 
an initially localized 
DF towards the selected self-similar DF and
find a {\it power-law} decay in time for
the corresponding (non-self-similar) perturbation. This decay is
much faster than 
the logarithmic decay
found for the
limiting solution \cite{LS1,LS2}. We will see that not only
the selected self-similar DF, but also the
decay exponent are determined solely 
by the analytical properties of $F(R,t=0)$ in the close vicinity 
of $R_{m}(t=0)$. Our
theoretical predictions show a very good agreement with simulations.

We will start with  
the asymptotic theory.  Solving the
problem analytically
is made possible by a change of variables that employs the compactness
of the support $[0, R_m (t)]$ of the DF. 
Introduce a scaled drop radius and a 
new time:
\begin{equation}
x(R,t) = \frac{R}{R_m(t)}\quad \mbox {and} \quad
\tau =  \int\limits_0^t  \frac{dt^\prime}{R_m^{n+2}(t^\prime)}\,,
\label{3}
\end{equation}
and a scaled DF
\begin{equation}
G(x,\tau)= R_m^4 [t(\tau)]\, F[R(x,\tau),t(\tau)]\,.
\label{V4}
\end{equation}
In the new variables
Eqs. (\ref{1}) and (\ref{2}) can be rewritten as
\begin{eqnarray}
(\partial G/\partial\tau)+
\left[v\, (x^{-n}-x) +x-x^{-n-1}\right]
(\partial G/\partial x) \nonumber\\
+\left[(n+1)x^{-n-2} - n v x^{-n-1}-4 (v-1) \right]\,  G =0
\label{BE1}
\end{eqnarray}
and $\int\limits_0^1 G(x, \tau) x^3\, dx = Q\,,$ respectively,
where $v(\tau)=R_m [t(\tau)]/R_c [t(\tau)]$. The 
function $G(x,\tau)$ is nonzero on the interval $0<x<1$ and zero elsewhere.

We will see in a moment that a self-similar solution for $F(R,t)$ 
corresponds to
a {\it steady-state} solution for $G(x,\tau)$. Therefore,
we are looking for the
solution in the following form: 
\begin{eqnarray}
G(x, \tau)  = \Phi_0(x) + \Phi_1(x)\, e^{q\tau}  + \dots\ ,\nonumber\\
v(\tau)  =  v_0 + v_1\,  e^{q\tau} + \dots\ , 
\label{B1}
\end{eqnarray}
where $q$ is a (sought for) complex number. Both 
$\Phi_0(x)$, and 
$\Phi_1(x)$ are localized
on the interval $[0,1]$. The perturbation 
must not change
the normalization condition (\ref{2}) and the asymptotics of
the unperturbed solution near the point $x=1$, that is,
$\int\limits_0^1 x^3\,\Phi_1 (x)\, dx =0$ and 
$\Phi_1  = {\cal O}(\Phi_0)$ at $x \to 1$.

A family of steady-state solutions $\Phi_0(x)$ 
(parameterized by $v_0$)
is obtained from the zero-order equation
\begin{eqnarray}
\left[v_0\, (x^{-n}-x) +x-x^{-n-1}\right]
(d \Phi_0 / d x) \nonumber \\
+\left[(n+1)x^{-n-2} - n v_0 x^{-n-1}-4 (v_0-1) \right]\, \Phi_0 =0\,.
\label{Z}
\end{eqnarray}
Integration of this equation in elementary functions is possible 
for $n=-1,0,1$ and $2$ (that is, for most cases of physical 
interest \cite{SS}). For 
example, for $n=0$
one has
\begin{equation}
\Phi_0(x)= C_Q\,x\, (1-x)^\alpha\, (x_2-x)^\nu\,,\,\,0\le x \le 1\,,
\label{UP1}
\end{equation}
where
\begin{equation}
\alpha   =  \frac{4v_0 -5}{2-v_0}\ ,\quad
\nu  = \frac{v_0-5}{2-v_0}\ ,\quad 
x_2   = \frac{1}{v_0-1}\ , 
\label{B3}
\end{equation}
while $C_Q$ is determined from the 
condition $\int\limits_0^1 x^3\, \Phi_0(x)\, dx$ \linebreak $=Q$.
This family of solutions 
is defined for $5/4 < v_0 < 2$. It corresponds to the
family of {\it self-similar} solutions for $F(R,\,t)$ obtained 
in Ref. \cite{MS96}.

For $n=1$ one obtains 
\begin{equation}
\Phi_0= C_Q x^2 (1-x)^\alpha (x-x_{-})^{\gamma_2-\gamma_1}
(x_{+}-x)^{-\gamma_1-\gamma_2}\,.
\label{UP10}
\end{equation}
Here
\begin{equation}
\alpha   =  \frac{5v_0 -6}{3-2v_0}\,,\quad
\gamma_1 = \frac{12-7v_0}{6-4v_0}\,,\quad 
\gamma_2  = \frac{3v_0}{(6-4v_0)s}\,, 
\label{B30}
\end{equation}
$x_{\pm}=(-1\pm s)/2$, $s=[(v_0+3)/v_0-1)]^{1/2}$ 
and $0\le x \le 1$. This family is defined for
$6/5<v_0<3/2$.

For any $n$, we will need to know the behavior of $\Phi_0 (x)$ and 
$\Phi_1 (x)$ in the close vicinity of $x=1$.
A simple analysis of Eq. (\ref{Z}) yields
$\Phi_0 = H_0 (\zeta) \,\zeta^{\alpha}\,$,
where $\zeta= 1-x$, $H_0(\zeta)$ is 
an analytic function on the interval $[0,1]$, 
$H_0(0)\neq 0$, and
\begin{equation}
\alpha=\frac{(n+4)\,v_0-n-5}{n+2-(n+1)\,v_0}\,.
\label{lam}
\end{equation}
The solution for $\Phi_0 (x)$ 
exists if $0<\alpha<\infty$, that is \linebreak
$(n+5)/(n+4)< v_0 < (n+2)/(n+1)$. This interval of permitted values
of $v_0$ is non-empty for any $n\ge-1$. [The case of $n=-1$ is the simplest:
$H_0 (\zeta)=const$.]

Now we go to the first order in Eq. (\ref{BE1}):
\begin{eqnarray}
\left[v_0\, (x^{-n}-x) +x-x^{-n-1}\right]
(d \Phi_1/d x) \nonumber \\
+\left[q-(n+1)x^{-n-2} - n v_0 x^{-n-1}-4 (v_0-1) \right]\, \Phi_1\nonumber \\
= v_1\left[(x-x^{-n})\, (d\Phi_0/dx) + (4+n x^{-n-1}) \Phi_0\right]\,.
\label{P2}
\end{eqnarray}
For a given $v_1$, this linear equation
can be solved in quadratures \cite{quadrature}. 
We will need 
only the 
leading asymptotics of the
solution in the close vicinity of $x=1$, so we write down the solution 
as
\begin{equation}
\Phi_1 = \zeta^\alpha\,\chi_1(\zeta)\, + \zeta^{\beta}\,\chi_2(\zeta) \,,
\label{phi}
\end{equation}
where $\beta=\alpha-q [n+2 - (n+1) v_0]^{-1}$, 
$\chi_1$ and $\chi_2$ are analytic functions 
on the interval $[0,1]$ and $\chi_{1,2}\, (0) \ne 0$. The solution 
exists if $Re\,{\beta}\ge 0$ which implies
$Re\, q < (n+4) v_0-n-5$. One can check {\it a posteriori} that 
this inequality holds.

Eq. (\ref{phi}) will be used later.
At this stage we notice that the still undetermined
``eigenvalue" $q$ must be selected by the initial condition.
To make this selection possible, we should exploit an (exact) 
invariance property of Eq. (\ref{1}). 
Consider the initial value problem $dR/dt=
V (R,t)\,,
R(0)=R_0$ that describes the characteristics 
of Eq. 
(\ref{1}).  If the solution of this problem, $R(t; R_0)$, 
is known, the
solution
of Eq. (\ref{1}) can be written in the following form: 
\begin{equation}
F(R,t)=F_0[R_0(R,t)]\, \partial R_0 (R,t)/\partial R,
\label{Seq5a}
\end{equation}
where $R_0(R,t)$ is the 
function inverse to $R(t, R_0)$ with respect to the 
argument $R_0$. Now, $R(t, R_0)$
is an analytic 
and monotonic function of $R_0$. Therefore, the inverse function 
$R_0(R,t)$ is also an analytic function of $R$, and so
$F(R,t)$  preserves
its analytic form along the characteristics $R=R(t; R_0)$, including the
``edge" characteristics $R_{m}(t)$.

We assume a power-law 
behavior of $F(R,\,t=0)$ in the close vicinity of $R=R_m(0)$.
More precisely, we assume that
\begin{equation}
F_0 (\xi) =\xi^{\lambda_1} g_1(\xi) + \xi^{\lambda_2} g_2(\xi)\,, 
\label{Seq8}
\end{equation}
where $\xi=R_{m}(0)-R>0$.  Here $\lambda_1$ and
$\lambda_2>\lambda_1$ are arbitrary positive numbers, such that
$\lambda_2-\lambda_1 \ne 1,\, 2,\, \dots\,$, and
$g_1(\xi)$ and $g_2(\xi)$
are analytic at
$\xi=0$ such that $g_{1,2}(0)\ne0$.
In view of
of the analyticity property mentioned above,
Eq. (\ref{Seq5a}) can be rewritten
as
\begin{equation}
F(R,t) = (\xi')^{\lambda_1} h_1(\xi',t) + (\xi')^{\lambda_2} h_2(\xi',t)\,,
\label{Seq8-1}
\end{equation}
where $\xi'=R_{m}(t)-R>0$, while
$h_1(\xi',t)$ and $h_2(\xi',t)$ are analytic functions of $\xi'$ at $\xi'=0$,
and $h_{1,2}(0,t)\ne 0$. 

Under the transformation (\ref{3}), (\ref{V4}) 
the variables $R$ and $F$ are multiplied by 
some quantities independent of 
$R$. Therefore, we can rewrite
Eqs.~(\ref{Seq8}) and (\ref{Seq8-1}) in the new 
variables $x$ and $\tau$ as follows. The initial DF
is now
\begin{equation}
G_0 (x)= \zeta^\lambda_1 \, g'_1(\zeta) +  \zeta^{\lambda_2} \, g'_2(\zeta)\,, 
\quad\quad \zeta>0,
\label{Seq8-2}
\end{equation}
where $g'_1(\zeta)$ and $g'_2(\zeta)$
are  analytic at
$\zeta=0$, $g'_{1,2}(0)\ne 0$, and we
remind that $\zeta=1-x$. 
Correspondingly, the time-dependent DF $G(x,\tau)$ 
can be
written as
\begin{equation}
G(x,\tau) = \zeta^{\lambda_1}\,h'_1(\zeta,\tau) + 
\zeta^{\lambda_2}\, h'_2(\zeta,\tau)\,,
\quad\quad \zeta>0,
\label{Seq8-3}
\end{equation}
where $h'_1(\zeta,\tau)$ and $h'_2(\zeta,\tau)$ are analytic functions of 
$\zeta$ in $\zeta=0$, and $h'_{1,2}(0,\tau)\ne 0$. 
 
The exponents $\lambda_1$ and $\lambda_2$, prescribed by the initial
conditions, remain invariant. 
Hence, the long-time asymptotics of Eq. (\ref{Seq8-3}) should coincide
with that given by Eqs. (\ref{B1}) and (\ref{phi}).   
A direct comparison 
yields $\alpha = \lambda_1$ and $\beta = \lambda_2$. The first equality
is nothing but a (weak) selection rule 
for the self-similar
solution, and the selected value of 
$v_0$ is
\begin{equation}
v_0=\frac{(n+2) \lambda_1+n+5}{(n+1)\lambda_1+n+4}\,.
\label{v0}
\end{equation}
The second equality determines $q$:
\begin{equation}
-q  = \frac{3 (\lambda_2-\lambda_1)}{(n+1)\lambda_1+n+4}\,.
\label{q}
\end{equation}
One can see that $-q$ is real 
and positive which means
stability. 
 
Going back to the ``physical" variables $R$ and $t$ is easy. Indeed, 
evaluating $R_m(t)$ on the self-similar solution, we obtain
$R_m (t) = \left[(n+2)\, (v_0-1) \,t\right]^{1/(n+2)}.$ Then, 
using 
Eq. (\ref{3}), we see that
$e^{q\tau}=t^{-\Gamma}$, a power-law decay in the ``physical" time. Here
$$\Gamma=\frac{3 (\lambda_2-\lambda_1)}{(n+2)(\lambda_1+1)} > 0\,.$$

If we limit ourselves to an important particular
case of a single ``non-trivial" exponent in the initial DF,
$G_0(x)=\zeta^{\lambda} g(\zeta)$, [where $g(\zeta)$
is analytic and $g(0)\ne 0$], then 
$G(x,\tau)=\zeta^{\lambda} h(\zeta,\tau)$,
where $h(\zeta,\tau)$ is an analytic functions of 
$\zeta$ and $h(0,\tau)\ne 0$.
Now, using  Eqs. (\ref{B1}) and (\ref{phi}), we obtain
\begin{equation}
-q  = 
\frac{3}{(n+1)\lambda+n+4}\,,
\label{qpart}
\end{equation}
unless the linear term in the Taylor series of $g(\zeta)$ at
$\zeta=0$ 
vanishes \cite{vanish}. This yields the power exponent
\begin{equation}
\Gamma=\frac{3}{(n+2)(\lambda+1)}\,.
\label{Qpart}
\end{equation}

In the limit of $\lambda\to
\infty\,$, we obtain $\Gamma \to 0$. 
Clearly, it corresponds to the
logarithmically slow
decay obtained for the limiting solution \cite{LS1,LS2}.

Therefore, both the self-similar DF, and the power-law decay 
rate
of a small perturbation around it are uniquely determined 
by the asymptotics of the initial DF in the close vicinity of the maximum
domain size,
$R=R_m(0)$.

We verified the theory (in the cases $n=0$ and $1$) by 
performing extensive numerical simulations with Eq. (\ref{1}) and
an explicit equation for $\dot{R_c}$ that follows from Eqs. (\ref{1}) and
(\ref{2}).  As the dynamics is 
extremely sensitive to small changes
in the vicinity of $R=R_m (t)$, we needed an algorithm that 
preserved the
compactness of the DF and kept a high accuracy near the edge point
$R=R_m (t)$. A 
simple and efficient Lagrangian algorithm was developed 
\cite{Giron} that 
satisfied these requirements. Typical simulation
results for 
the interface-controlled OR, $n=0$, are presented in Figs. 1 and 2. 
Figure 1 
shows convergence of an initially
localized DF, $F(R,0)= R\,(5-R)^{\lambda}$ with $\lambda=1$,
towards the selected self-similar DF 
(\ref{UP1}), for which Eq. (\ref{v0}) predicts $v_0=7/5$. The inset shows
convergence of $v(t)$ towards 
$v_0=7/5$. The convergence exponent $\Gamma_{exp} = 0.76$ found numerically 
agrees very well with 
our theoretical prediction $\Gamma_{th} = 0.75$. Figure 2 shows the convergence
exponents $\Gamma_{exp}$ found numerically for 
different $\lambda$. A good 
agreement with the
theoretical
curve $\Gamma_{th}=3/[2(\lambda+1)]$ is seen. We also observed a good agreement
between the theory and simulations in the case of 
the diffusion-controlled OR, $n=1$.

\begin{figure}[h]
\vspace{0.0cm}
\hspace{0.0cm}
\rightline{ \epsfxsize = 9.0cm \epsffile{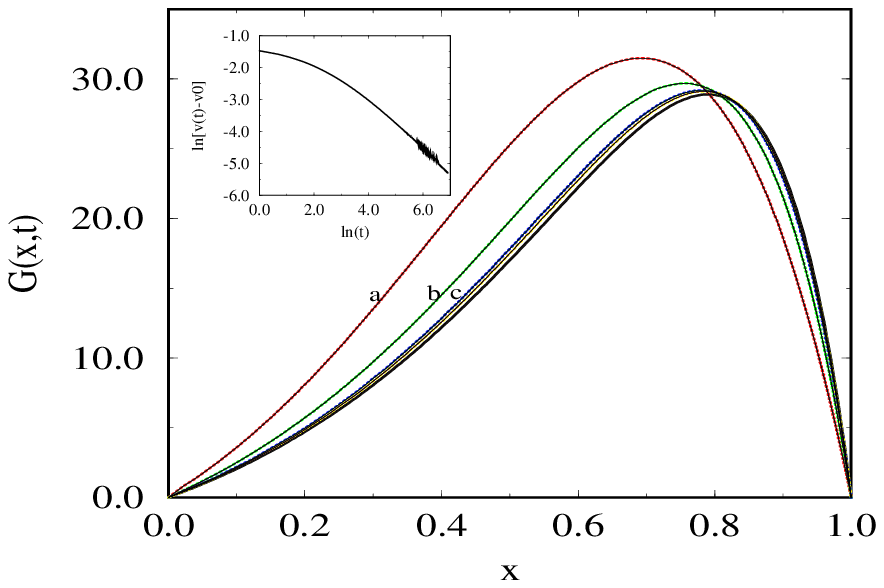}}
\caption{
Convergence of an initially localized DF with $\lambda=1$ 
towards the selected self-similar
DF (\ref{UP1}) with $v_0=7/5$ (solid line). Numerical solutions are shown
by dotted lines at time moments
$t = 20$ (a), $100$ (b), $500$ (c) and 1000. The inset shows the 
convergence of $v (t)$ 
towards $v_0=7/5$.
\label{Fig. 1}}
\end{figure}

We have demonstrated  
that only a weak selection is possible in the ``classical" model of OR. 
To get a 
{\it strong} selection rule, one obviously must go beyond the ``classical" 
model.

This work was supported in part by a grant from Israel Science Foundation,
administered by the Israel Academy of Sciences and Humanities, and by the
Russian Foundation for Basic Research (grant No. 96-01-01876).

\begin{figure}[h]
\vspace{0.0cm}
\hspace{-1.5cm}
\rightline{ \epsfxsize = 7.0cm \epsffile{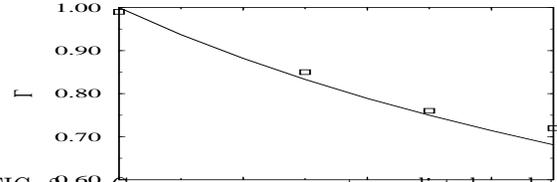}}
\caption{
Convergence exponents predicted analytically (line) and 
found numerically (squares) 
for different $\lambda$.
\label{Fig. 2}}
\end{figure}


\begin{references}
\bibitem{quadrature} The solution is obtained by making 
the ansatz $\Phi_1(x)=\Phi_0(x)\,f(x)$ and directly solving the 
equation for $f(x)$.
\bibitem{vanish} If it does vanish, an additional
factor 2 appears in the numerators of
the right hand sides of Eqs.~(\ref{qpart}) and (\ref{Qpart}). In 
general, if $k$ is
the order of the first non-vanishing term in the Taylor series, an additional
factor $k$ appears.
\end{references}
\end{document}